## Tautomerism in liquid 1,2,3-triazole: a combined Energy-Dispersive X-Ray Diffraction, Molecular Dynamics and FTIR study

--Manuscript Draft--

| | |
|---|---|
| **Manuscript Number:** | STUC-D-13-00015 |
| **Full Title:** | Tautomerism in liquid 1,2,3-triazole: a combined Energy-Dispersive X-Ray Diffraction, Molecular Dynamics and FTIR study |
| **Article Type:** | SI: Domenicano Special Collection |
| **Keywords:** | 1,2,3-triazole liquid phase, tautomers, X-ray diffraction, Molecular Dynamics and Density Functional Calculations |
| **Corresponding Author:** | Lorenzo Gontrani<br>CNR - ISM<br>Roma, ITALY |
| **Corresponding Author Secondary Information:** | |
| **Corresponding Author's Institution:** | CNR - ISM |
| **Corresponding Author's Secondary Institution:** | |
| **First Author:** | Marco Bellagamba, Dr. |
| **First Author Secondary Information:** | |
| **Order of Authors:** | Marco Bellagamba, Dr. |
| | Luigi Bencivenni, Dr. |
| | Lorenzo Gontrani |
| | Leonardo Guidoni, Prof. |
| | Claudia Sadun, Prof. |
| **Order of Authors Secondary Information:** | |
| **Abstract:** | In this work, we report a multitechnique (energy-dispersive X-Ray diffraction, computational methods and FT-IR spectroscopy) study of the tautomeric equilibrium of 1,2,3-triazole, one of the few small nitrogen-containing eterocycles liquid at room temperature. The T-2H form (C2v symmetry) is found to be strongly favored in gas and solid phases, whereas the neat liquid gives diffraction patterns that can be interpreted satisfactorily with the structure functions calculated from some molecular dynamics results for both T-2H and T-1H tautomers, although the T-2H form gives a slightly better agreement. |

# Tautomerism in liquid *1,2,3*-triazole: a combined Energy-Dispersive X-Ray Diffraction, Molecular Dynamics and FTIR study

Marco Bellagamba[a], Luigi Bencivenni[a], Lorenzo Gontrani[b], Leonardo Guidoni[c] and Claudia Sadun[a]

*This study is dedicated to Prof. Aldo Domenicano on the occasion of his 75$^{th}$ birthday.*

a) Dipartimento di Chimica, Università di Roma "La Sapienza", P.le Aldo Moro 5, 00185 Roma, Italy

b) CNR-Istituto di Struttura della Materia, Area della Ricerca di Roma Tor Vergata, Via del Fosso del Cavaliere 100, I-00133 Roma, Italy

c) Dipartimento di Scienze Fisiche e Chimiche, Università degli Studi de L'Aquila, Via Vetoio 2, I-67100 Coppito, L'Aquila, Italy

**Abstract**

In this work, we report a multitechnique (energy-dispersive X-Ray diffraction, computational methods and FT-IR spectroscopy) study of the tautomeric equilibrium of *1,2,3*-triazole, one of the few small nitrogen-containing eterocycles liquid at room temperature. The *T-2H* form ($C_{2v}$ symmetry) is found to be strongly favored in gas and solid phases, whereas the neat liquid gives diffraction patterns that can be interpreted satisfactorily with the structure functions calculated from some molecular dynamics results for both *T-2H* and *T-1H* tautomers, although the *T-2H* form gives a slightly better agreement.

**Author for correspondence**

Lorenzo Gontrani, CNR-ISM; lorenzo.gontrani@gmail.com

## Introduction

Triazacyclopentadiene ($C_2H_3N_3$), better known as *1,2,3*-triazole (CAS 288-36-8; CHEBI 35566) is an aromatic heterocyclic molecule which may exists in two tautomeric forms, which differ by the position of the hydrogen atom linked to the nitrogen atom in the five-membered ring, namely *1H-1,2,3*-triazole and *2H-1,2,3*-triazole, from now on denoted in the paper as *T-1H* and *T-2H*, respectively (see the attached chart).

*T-2H*

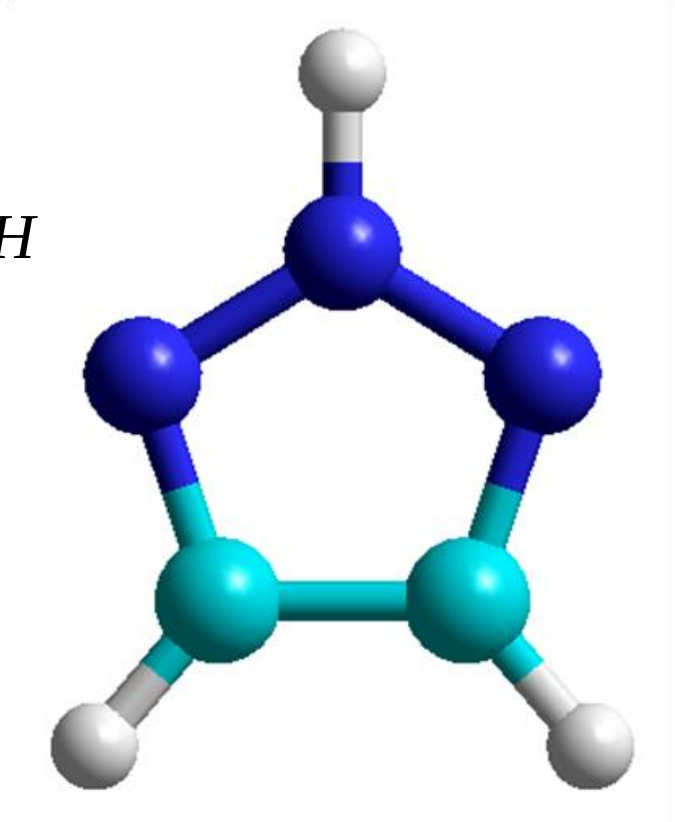

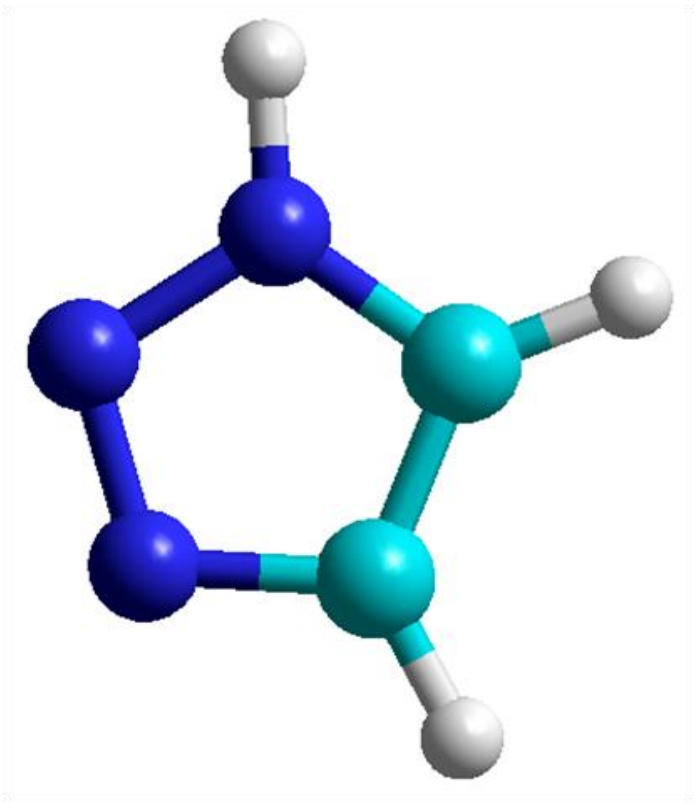

*T-1H*

*1,2,3*-triazole is a room-temperature stable liquid (melting point 296 K), while other heterocyclic substances having contiguous nitrogen atoms (pyrazole, *1,2,4*-triazole or tetrazole) are solid. It is a synthetic molecule and its derivatives are largely employed as building blocks for several complex chemical products in pharmacology [1] and food-farming [2].

*1,2,3*-triazole solid phase was studied by Goddard *et al.* [3] by single crystal X-ray diffraction. The crystal phase was obtained from *in situ* crystallization [4] of the liquid at 238 K and it was found to be composed of a mixture of both tautomers in 1:1 molecular ratio. This X-ray study reveals that in the crystal structure, the two tautomers have virtually identical bond distances, while bond angles differ slightly between each other.

Gas-phase tautomerism was evaluated by Begtrup *et al.* [4] by microwave spectroscopy, electron diffraction and *ab initio* calculations. Microwave spectroscopy indicates that *1,2,3*-triazole gas-phase consists of the *T-2H* tautomer, being the estimated *T-1H:T-2H* ratio quite low ($10^{-3}$ at room temperature). *Ab initio* calculations [4,5] confirm that the *T-2H* tautomer is between 15 and 20 kJ $mol^{-1}$ more stable than the *T-1H* species, although the free energy difference between the two forms is strongly dependent on the theoretical level of the calculation. On the contrary, liquid phase tautomerism is still an open matter. $^{15}N$-NMR studies [5] accomplished at room temperature for some methyl derivatives of the 1H and 2H tautomers in $DCCl_3$ and $(CH_3)_2SO$-d6 suggest that the *T-2H* population is 35% in $DCCl_3$ and 55 % of the total in $(CH_3)_2SO$-d6; a similar figure (40%) was determined by Begtrup from $^{13}C$-$^{1}H$ coupling studies [6]. Wofford [7] explained these results claiming solvent polarity and hydrogen bonding; according to this interpretation, the tautomer able to form the strongest hydrogen bond with a polar solvent is the most preferred molecular species. Actually, the electric dipole moment is expected to be larger for the *T-2H* tautomer than for the *T-1H* one, because the polarity of the *NH* bond is enhanced in the former for the presence of two contiguous nitrogen atoms. A result in complete disagreement with the results of Wofford [7] was obtained by Lunazzi *et al*. [8], who investigated the 0.05 M solution of *1,2,3*-triazole in chloroform by $^{1}$H-NMR, reaching the conclusion that at room temperature the solution consists of 80% *T-2H* tautomers. Likely, the tautomeric equilibrium depends on concentration; moreover, Wofford obtained his results indirectly from *1,2,3*-derivatives [7]. Lunazzi *et al.* [8] were also able to determine the dependence of the tautomeric population on solvent polarity. For instance, the *T-2H:T-1H* population for 0.01 M solutions of *1,2,3*-triazole in toluene at 300 K is 4:1, while in chloroform at 175 K the same population ratio is reversed. At last, Fabre *et al.* [9], measuring dipole moment of *1,2,3*-triazole in benzene, found that in this solvent 83% of the solute exists in the *T-2H* form. Although the matter has not yet been studied in detail for aqueous solutions, literature data agree with the fact that the tautomeric equilibrium would always seem to be favorable to the *T-2H* form of the molecule [8, 10].

Tautomerism of *1,2,3*-triazole is a valuable source for interpreting and predicting chemical reactivity and molecular recognition and the characterization of tautomeric equilibrium is of great importance in biochemistry and pharmacology. For instance, cases of spontaneous mutation processes are correlated to the appearance of nitrogen bases of less abundant tautomers of a given molecule [11]. *1,2,3*-triazole is isoelectronic with *1,2,4*-triazole and its substitution in some purinic species is a reason prompting these studies. Although tautomerism in aqueous solution exerts the greatest interest, as biochemical processes occur in water environment, we have decided to first clarify tautomerism of liquid *1,2,3*-triazole, never studied so far under this aspect. Energy Dispersive X-Ray Diffraction (EDXD) measurements and theoretical quantum-mechanical and molecular dynamics studies were undertaken for the purpose. The combined use of computational methods and experimental techniques, in particular a non-conventional type of X-Ray diffraction, has already proved to be a powerful tool to study disordered systems such as liquid and amorphous compounds and would guarantee more reliable system modeling [12-15]. Although the weak spot of X-Ray diffraction methods is the inability of precisely locating hydrogen atoms, and the two tautomers of *1,2,3*-triazole just differ for the position of *NH* bond in the molecular ring, the presence of this polar group produces, through intermolecular hydrogen bonding, different supramolecular structures depending on which tautomer is involved in liquid phase organization.

Effective role of computational methods is particularly valuable to interpret the diffraction results and to obtain reliable structural information for this liquid.

## Experimental section

All the compounds were purchased from Aldrich, and were of analytical-grade purity.

### Energy dispersion X-Ray diffraction (EDXD)

EDXD measurements were performed using the non-commercial energy-scanning diffractometer projected and assembled in the Department of Chemistry of Rome "La Sapienza" University. Detailed description of both instrument and technique can be found in previous papers by our group (e. g. Ref. 12). The experimental protocol (instrument geometry and scattering angles) of the data acquisition phase is analogous to that followed in previous experiments. The appropriate measuring time (i.e. number of counts) was chosen as to obtain scattering variable (Q) spectra with high signal to noise ratio (500000 counts on average). The expression of Q is

$$Q = \frac{4\pi \sin\vartheta}{\lambda} \approx 1.0136 E \sin\vartheta$$

where E is expressed in keV and Q in $Å^{-1}$, $2\vartheta$ being the scattering angle. The various angular data were processed according to the procedure described in literature, normalized to a stoichiometric unit of volume containing one triazole molecule and combined to yield the total static structure function, I(Q), which is equal to $I(Q) = I_{e.u.} - \sum_{i=1}^{n} x_i f_i^2$ where $f_i$ are the atomic scattering factors, $x_i$ are the number concentrations of the *i*-type atoms in the stoichiometric unit and $I_{e.u.}$ is the observed intensity in electron units ($electrons^2$).

Such function depends on the scattering contributions of all the particles of the system, according to the pairwise distances between them (in the case of X-Ray, the atoms scatter the radiation through the electron clouds surrounding them). Therefore, the structure function gives a sort of mathematical picture of the spatial disposition of the sample atoms. Such spatial arrangement can be figured out by confronting the experimental structure function with those calculated from model systems. The theoretical structure function for all the geometrical models built throughout our study was obtained using the Debye function [16,17]. This equation states that the structure function for a system made up of k particles located at distance r is

$$I_{Debye}(Q) = \sum_{m>n} x_m x_n f_m f_n \frac{\sin(r_{mn} Q)}{r_{mn} Q}$$

where $r_{mn}$ is the distance between particles m and n, f and x their scattering factors and concentrations, respectively. The function was then multiplied by Q and by a Q-dependent sharpening factor, M(Q), having chosen oxygen as the "sharpening atom".

$$M(Q) = \frac{f_O^{\,2}(Q)}{f_O^{\,2}(0)} \exp(-0.01Q^2)$$

The actual functions used in all the rest of the discussion are, therefore, QI(Q)M(Q).

Fourier transformation of the structure function leads to the total radial distribution function (RDF) in distance-space, which is expressed with slightly different formulations [18,19], according to the spatial range considered: the "total pair correlation function" G (r), most appropriate for intramolecular (short-range) contacts

$$G(r) = 1 + \frac{1}{2\pi^2 \rho_0} \int_o^{Q\max} Q\, I(Q) M(Q) \sin(rQ)\, dQ$$

$$Diff(r) = \frac{2r}{\pi} \int_o^{Q\max} Q\, I(Q) M(Q) \sin(rQ)\, dQ$$

and the Diff(r) function, that is the most suited to point out intermolecular (medium-long range) interactions. We used the value of 20 $Å^{-1}$ (*i.e.* the largest *Q* value measured) as the upper limit of integration. Therefore, the model validity was assessed by confronting model structure functions and radial distribution functions with the experimental counterpart.

Particular attention was devoted to the elimination of spurious ripples that appear in the experimental radial pattern at low r values (usually below 1 Å), that was accomplished through the comparison with the calculated single molecule Diff(r) (from the model structure, see below) and subsequent Fourier anti-transformation into a "ripple-free" I(Q).

## Infrared Spectra

FTIR infrared spectra of *1,2,3*-triazole were measured in different conditions, that is in liquid phase and by trapping vapor phase in equilibrium with the liquid phase in argon matrix at 12 K. The spectra of liquid sample (Sigma-Aldrich, 98%) were recorded using reflectance infrared Fourier transform technique by placing the sample inside a suitable cell provided with CsI optical windows. Matrix isolation technique was employed to record low temperature FTIR spectra in argon matrix at 12 K. Details of the experimental assembly are given elsewhere [20]. A Bruker IFS 113-v interferometer was used for all the measurements.

Liquid sample vaporization under high vacuum was accomplished in the temperature range 315-335 K. High purity argon (99.98%), was used as matrix gas. The matrix gas and sample vapor were trapped onto the surface of a highly reflecting gold-coated copper block, kept at 12 K during co-deposition and scanning. The spectra were recorded in the 4000 – 400 $cm^{-1}$ range cumulating 200 scans with a resolution of 1 $cm^{-1}$. The same equipment and instrumentation was used for studying the FTIR spectrum of extremely thin molecular layers obtained from condensation of vapor phase at cryogenic temperature onto the surface of the gold-coated copper block using He (99.98%) as carrier gas.

## Computational Details

The calculations were accomplished using a Linux Opteron based cluster, composed of HP DL servers in different hardware configurations. Geometry optimizations, harmonic frequencies and single energy point calculations were performed using the Gaussian 03 package [21]. Molecular dynamics calculations were carried out using the GROMACS package (version 3.3.2) [22]. Two different simulations were carried out, one for each tautomer. In the model, 3021 molecules (*T-1H* or *T-2H*) were placed inside a cubic box of 66 Å edge length (volume 287500 $Å^3$). The edge chosen was slightly larger than twice the value of the last distance peak (of neat shape) observed in experimental radial distribution function (30 Å). The calculation of a distance-dependent quantity under cubic periodic boundary conditions (applied in the simulation, see next), in fact, requires the cell edge to be at least two times larger than the maximum value of the distance [23].

The energy of the system was modelled with the two-body Generalized Amber Force Field (GAFF), with 8.0 Å cut-off and standard PME treatment of electrostatic interactions under Periodic Boundary Conditions (PBC). RESP atomic charges were used [24]. The simulation was accomplished through four distinct steps: 1) 200 steps of Conjugate Gradient minimization 2) A short NVT dynamics with distance restraints 3) Thermal equilibration at 300 K and 1 bar (~1100 *ps*) in NPT *ensemble* 4) Production run (700 ps). The actual reaching of equilibrium was checked by monitoring the trend of energy, density and of the calculated structure function (on a single frame) *vs* simulation time. The system was regarded as equilibrated, when the fluctuations of the density and of the energy were less than 1%. Furthermore, the difference between structure functions of different trajectory snapshots became negligible, and the only differences could be found in the small peaks oscillating around the average curve. The densities of the simulated liquid were 1191 kg $m^{-3}$ (*T-1H*) and 1187 kg $m^{-3}$ (*T-2H*) in agreement with the experimentally measured value 1192 kg $m^{-3}$.

## Results and discussion

### Energy Dispersion X-ray Diffraction Study

To assess the tautomerism of *1,2,3*-triazole in liquid phase, EDXD diffraction measurements were carried out, as described in the experimental section. The measured functions are reported in Fig. 1 (structure function) and Fig 2 (radial distribution function).

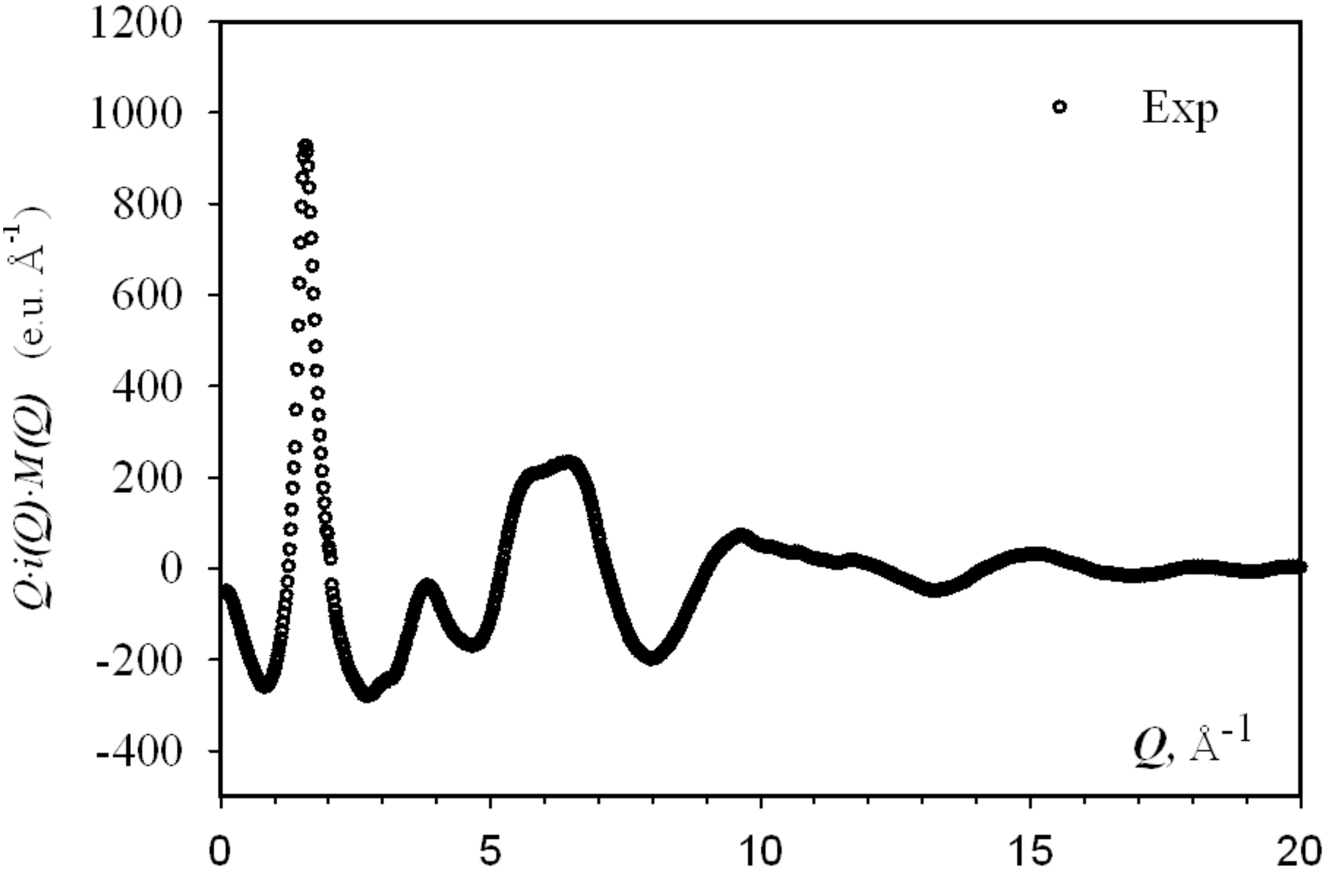

**Fig. 1** ***1,2,3*-triazole experimental structure function I(Q)**

No sharp peak is present in the radial distribution function that shows nothing but the typical damped oscillations of liquids. At low Q values a sharp and high intensity peak suggests a quite ordered character of this liquid because molecules interact with each other even at large distance (a few tens of Å), involving several correlation spheres. Actually, the presence of donor and acceptor sites in *1,2,3*-triazole and the consequent formation of a hydrogen bond network strongly favor high intermolecular organization in liquid phase. The first two sharp peaks of the radial distribution function represented in Fig. 2 are all related to the molecular structure of *1,2,3*-triazole and are followed by marked in-phase damping oscillations for increasing r values, revealing the molecular organization of the liquid. Oscillations occurring beyond 25 Å are no longer in-phase and are quite noisy, indicating that the structural correlation is lost at large distances. Evidently, single molecule contributions to the radial distribution function cause most of the oscillations of the structure function (see Fig. 3) and just the region at low exchanged momentum values is entirely due to intermolecular interaction. The weight of single molecule contributions to the total structure function was evaluated by calculating the single molecule Debye functions, which include intramolecular distances only, employing the geometries of both *T-1H* and *T-2H* tautomers, optimized at the B3LYP and MP2 levels with the aug-cc-pvtz and 6-311++G(3df,3pd) basis sets (both calculations provide the same high quality results and produce almost identical Debye functions). The experimental and calculated single molecule structure functions are shown in Fig. 3. It can be seen how the two theoretical curves nicely reproduce the experimental function at high Q values. The contribution of long-distance interactions (high r values), to structure function, in fact, damps much more quickly than the corresponding short distances term, and is negligible beyond 8 $Å^{-1}$. The two tautomers nearly show the same calculated structure function. This fact is not surprising, if the low sensitivity of X-ray diffraction towards light atoms and to the close structure similarity between *T-1H* and *T-2H* tautomers is considered. Though, some small differences are noticeable in the high-Q area of Fig. 3, which seem to support the idea that *T-2H* is the most favored structure. Such small differences can also be appreciated in the respective radial distribution functions (D(r) form) shown in Fig. 4. The peak present in the radial distribution function at 1.3 Å – 1.4 Å is due to adjacent atoms while the peak at 2.0 Å – 2.3 Å is the signal due to interaction between atoms separated by two bonds. Both peaks occur at the same r values in the radial distribution function of the two tautomers, although in the second peak the intensity of the experimental pattern is reproduced more precisely by the *T-2H* tautomer. Actually, *T-1H* and *T-2H* tautomers have sufficiently different bond angles, so that the values and relative weight of larger and shorter distances between atoms across the ring leads to different peak shapes. 1,3 contacts between heavy atoms in the two forms are reported in table 1.

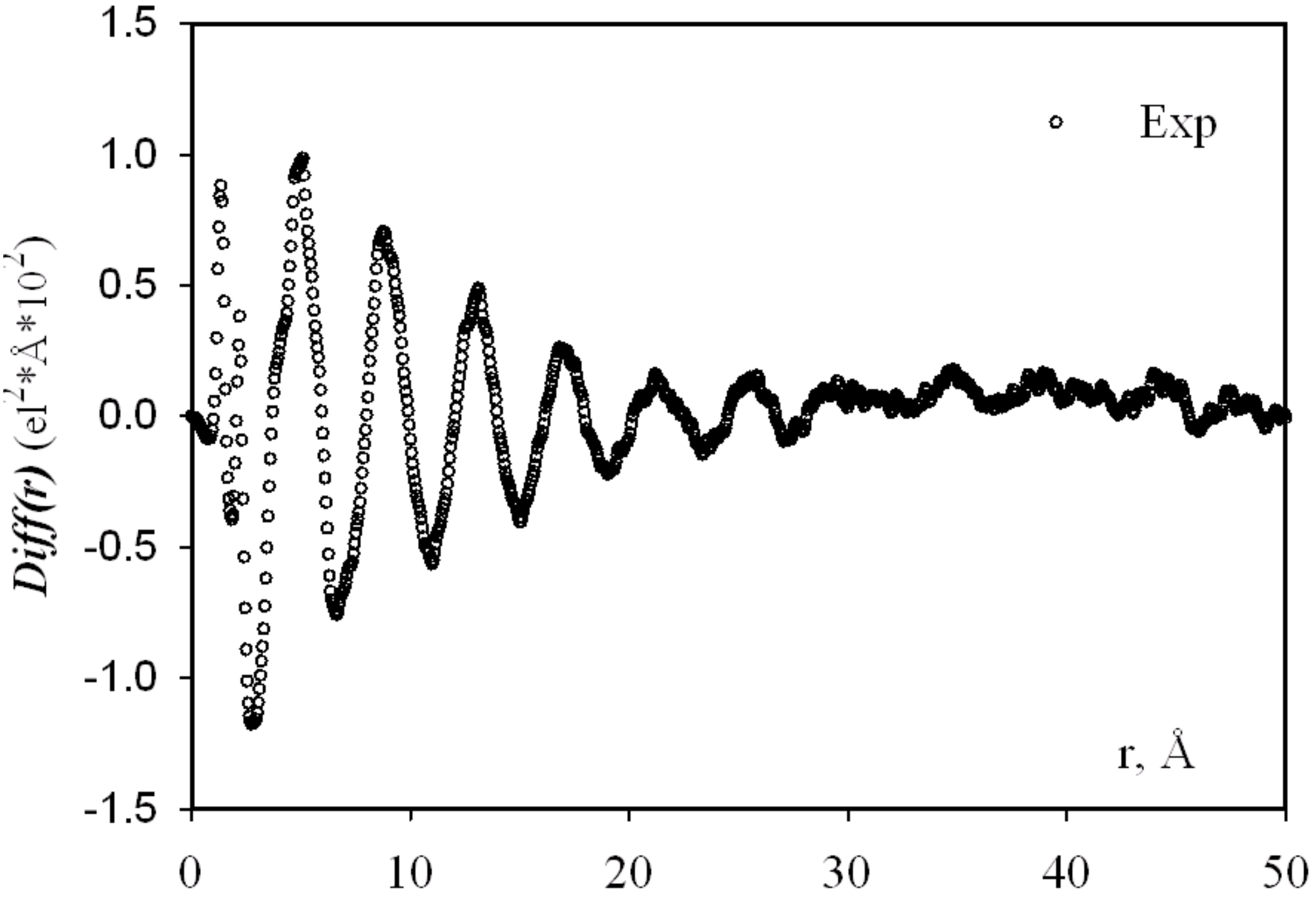

**Fig. 2** ***1,2,3*-triazole experimental radial distribution function Diff(r)**

Table 1 – 1,3 contacts in T-1H and T-2H forms (*ab initio* structures)

| **Pair** | **T-1H** | **T-2H** |
|---|---|---|
| N1-N3 | 2.15 | 2.31 |
| N1-C3 | 2.19 | 2.25 |
| N2-C2 | 2.20 | 2.10 |
| N2-C3 | 2.24 | 2.10 |
| N3-C2 | 2.16 | 2.25 |

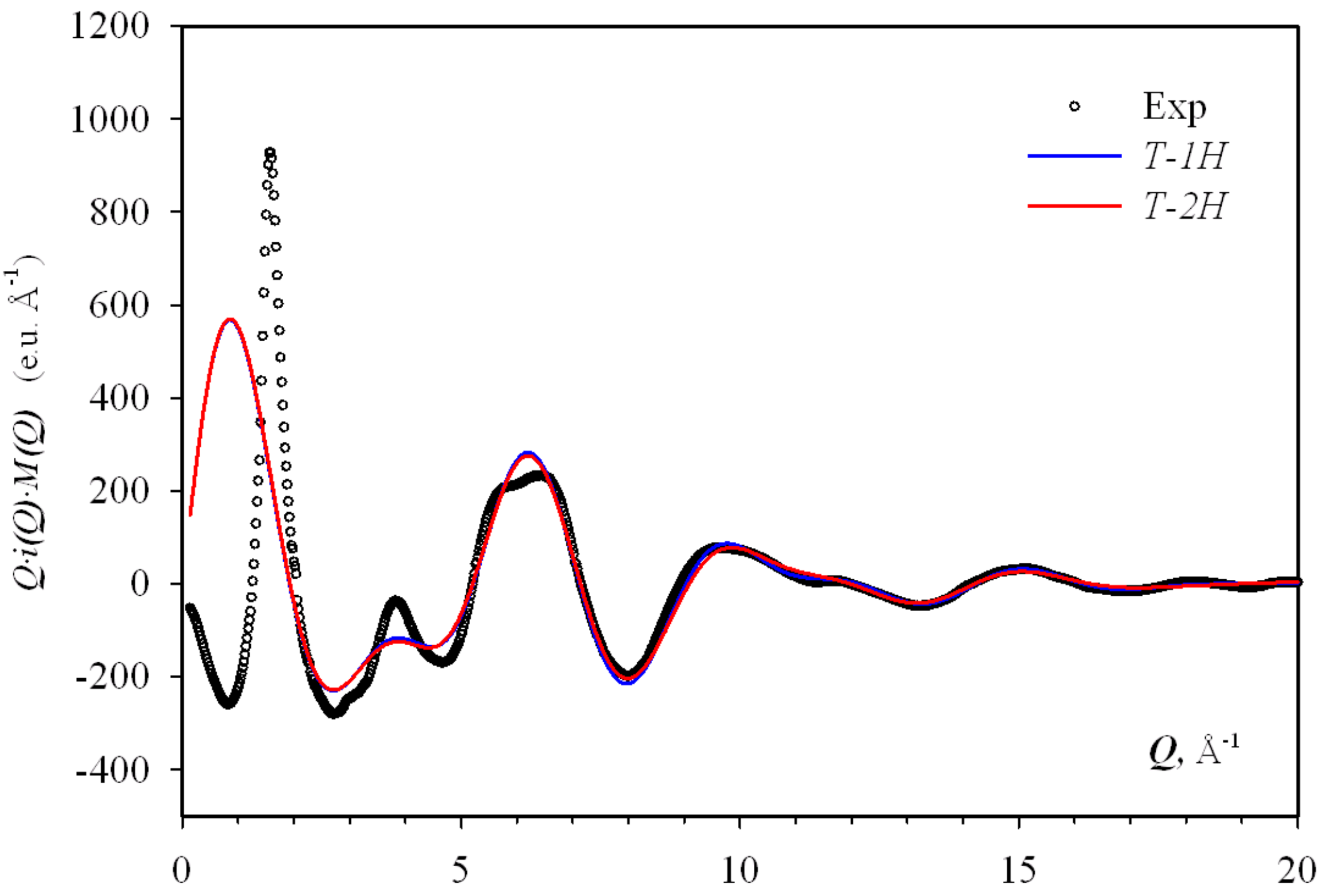

**Fig. 3 Comparison of T-1H and T-2H single molecule structure functions**

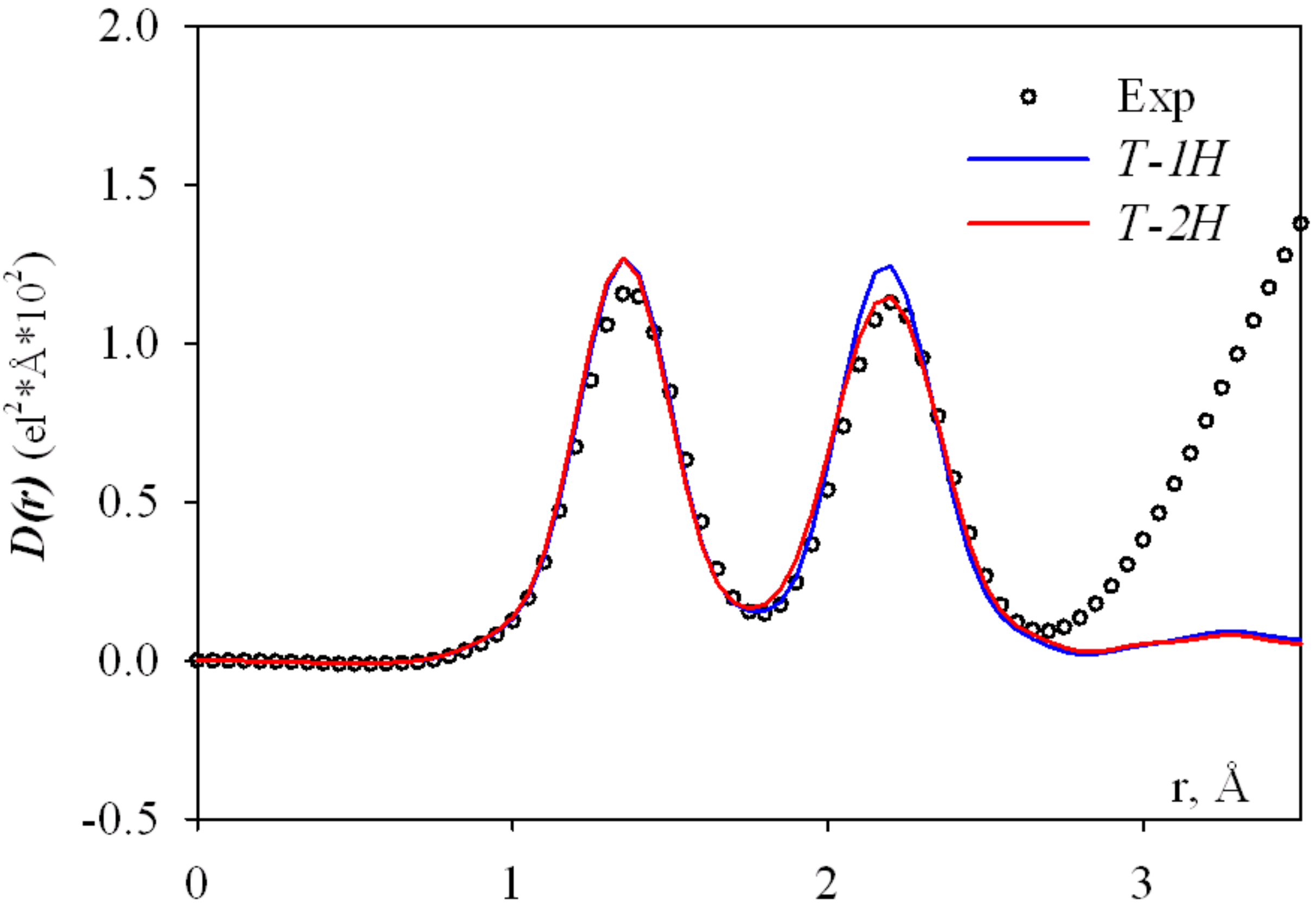

**Fig. 4 Comparison of T-1H and T-2H single molecule radial distribution functions D(r)**

The broad peaks falling beyond the intramolecular peaks (see Fig. 2) refer to all the intermolecular correlations of the systems, and their reproduction is hardly feasible with a single structural model without imposing large thermal factors to the interatomic distances, with the aim to reproduce the structural variability typical of a liquid. A more correct approach, in the framework of statistical mechanics, is to perform an average of all the instantaneous microstates (configurations) of a thermodynamic *ensemble*, representative of the real system. In this work, such *ensemble* average was performed by calculating the structure function as well as the radial distribution function for the frames of a trajectory obtained by a Molecular Dynamics simulation.

**Molecular Dynamics Study**

The comparison between EDXD and MD data was performed first for *T-1H*, using five different models, namely: for a given frame, chosen at random in the last 500 ps of the trajectory, a sphere containing 300 molecules; a sphere containing 500 molecules (model "1a") the whole cell (model "1box"), one "extended" cell, made up of two cells aligned along one cartesian axis (model "2box"); all the frames of the productive trajectory ("average").

The calculations allowed us to evaluate the model size effect in reproducing the experimental spatial correlation that extends up to about 30 Å in this system, according to the radial distribution function we obtained from our measurements. In the spherical models, to be sure that the result did not depend on the particular choice of the cell center, several positions in the cell were considered, and the result was found to be pretty insensitive to the specific sphere center, indicating that the system was homogenous. From the plots reported in Fig. 5 it appears that the size effect is apparently small in the reciprocal space (I(q)), and is limited to the first peak (1.58 $Å^{-1}$ – Fig. 5 inset), which is satisfactorily reproduced only if the entire cell is taken into account (model "1 box"). This fact is not surprising, considering that the first peak is originated by medium-long range (intermolecular) contacts mostly, and in the spherical models, although the radius is more than 20 Å, the number of molecules lying far apart is only a small portion of the total.

For the complete cell model structure function, the agreement is very good in the entire Q range, and the use of a more extended model (made of two cells along a single coordinate axis), does not improve the first peak significantly (Fig. 5). This view is supported by the analysis of radial distribution functions (Fig. 6). Whereas only the first two intermolecular peaks (peaks 3 and 4) after the intramolecular ones (1 and 2) are reproduced with a small model, whose spatial uncertainty (loss of correlation) is already evident at 10 Å, it is only with the complete box (models "1 box" and "2 box") that the Diff(r) oscillations are correctly reproduced. The remarkable noise in the single-frame functions ("1 box") is due to the already discussed structural variability. If all the frames of the productive trajectory are used, the long-range noise is integrated out, and the agreement of the curve ("*average*") is very good up to at least 30 Å (Fig. 7).

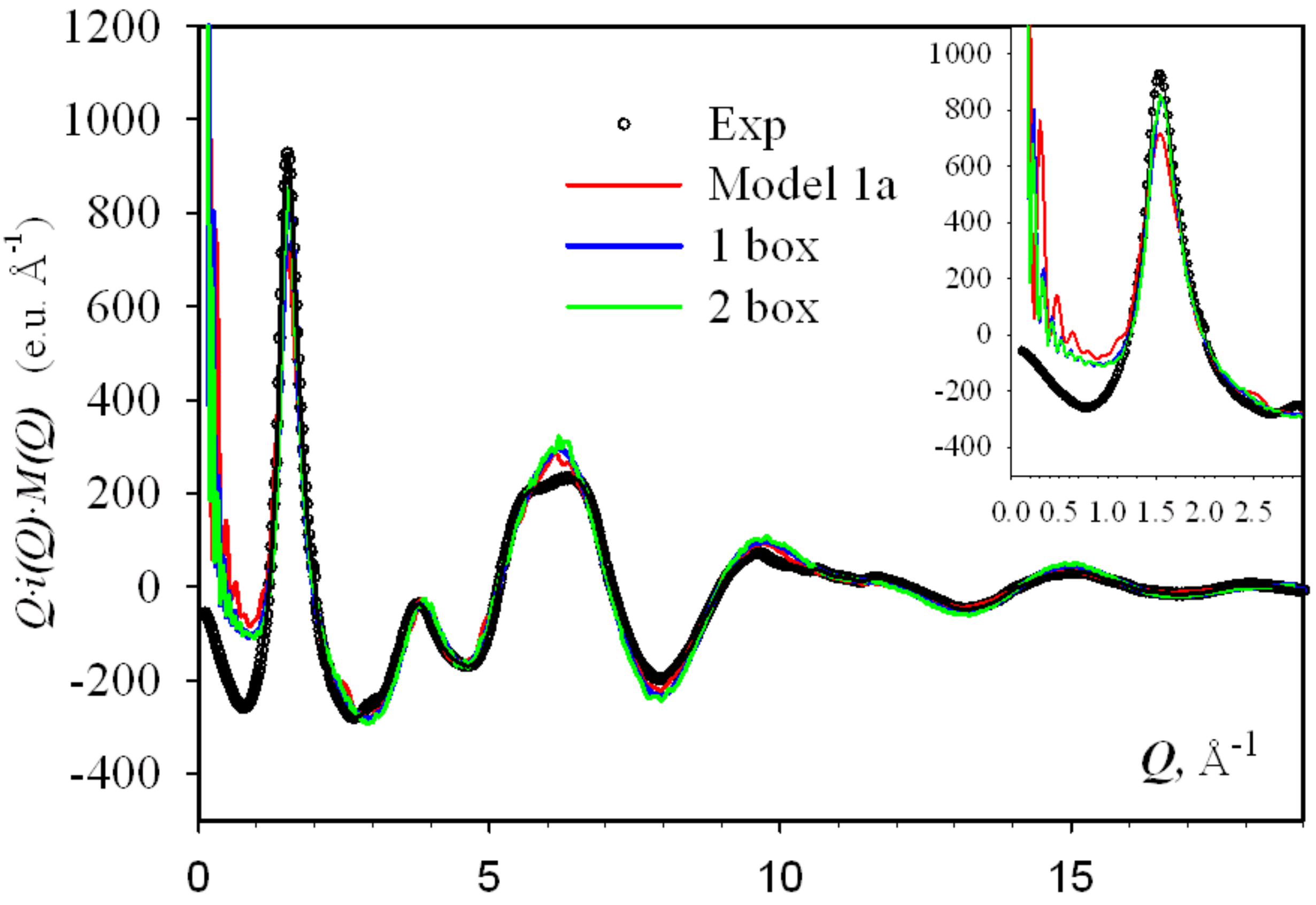

**Fig. 5 Different size models I(Q) *vs* experimental pattern**

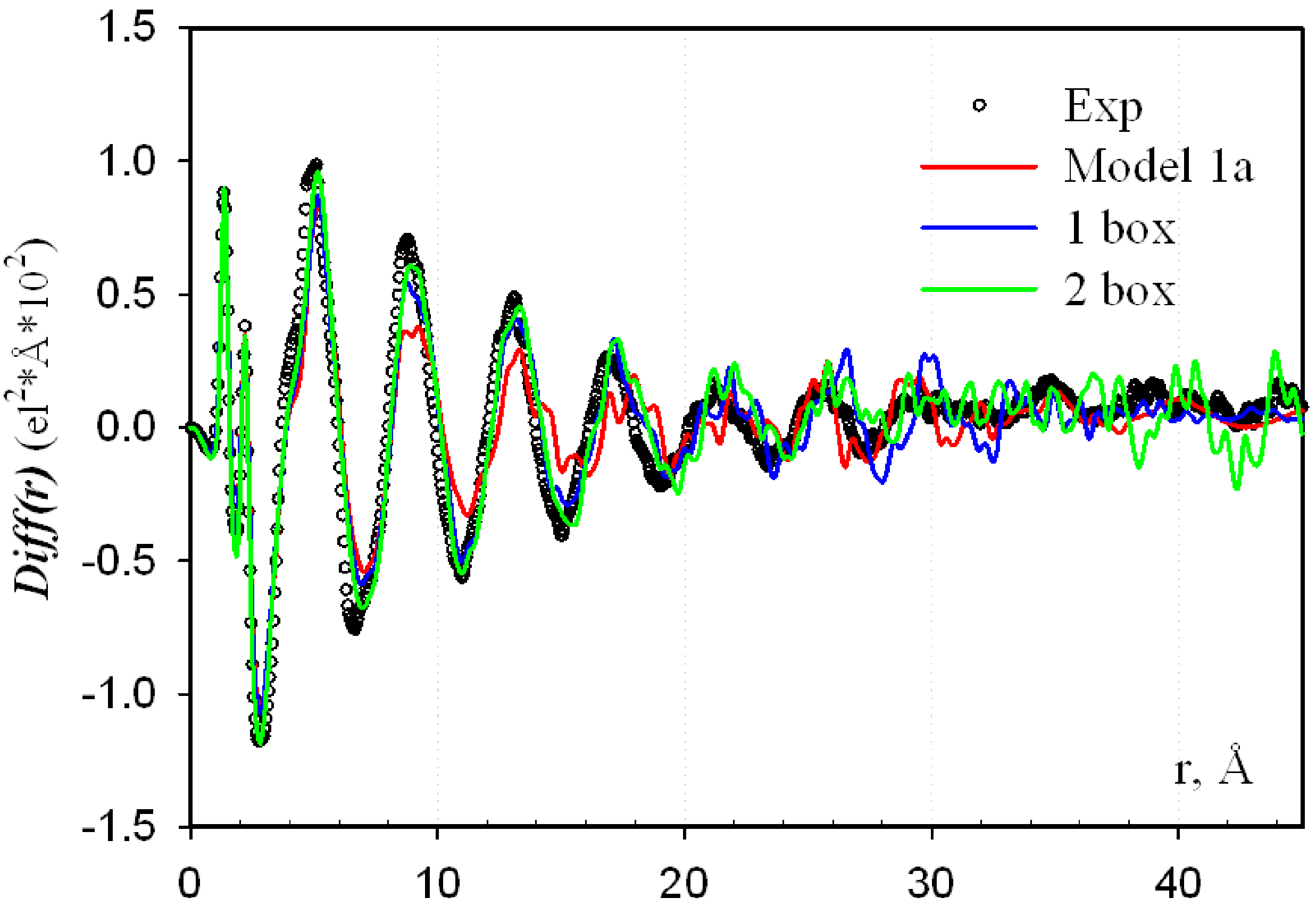

**Fig. 6 Different size models Diff(r) *vs* experimental pattern**

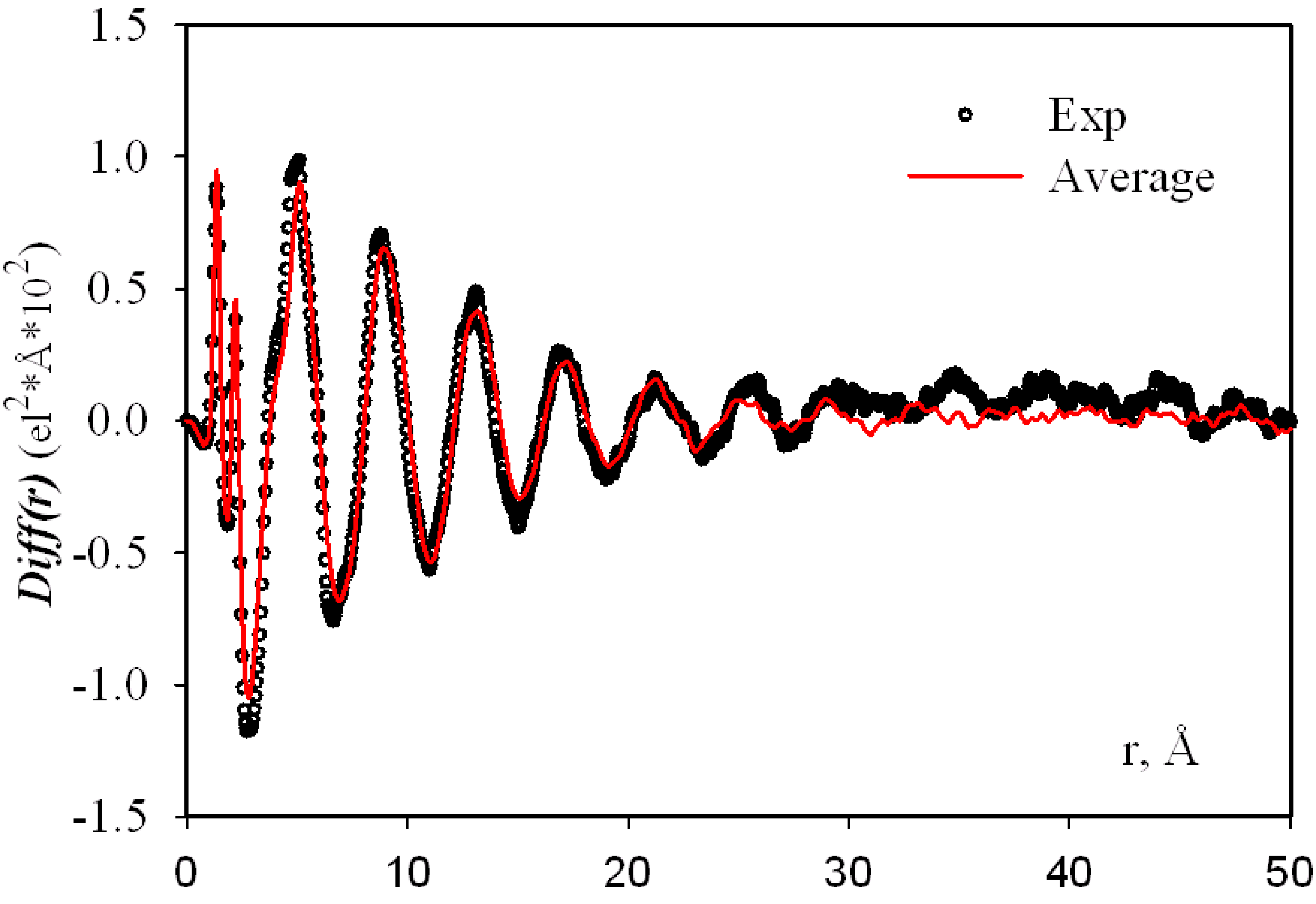

**Fig. 7 MD average Diff(r) *vs* experimental pattern**

**_T-1H / T-2H_ X-Ray patterns comparison**

The approach just described for *1H-1,2,3*-triazole was used for the analysis of the other tautomer (*T-2H*) and for the 1:1 mixture (MIX), composed of 1510 molecules of either form.. The overall comparison of the three models suggests that all reproduce the oscillations of the experimental curve satisfactorily, though *T-2H* model best fits experimental data in both I(Q) and Diff(r) curves. In particular, besides the better reproduction of the second molecular peak, already discussed using the static models (Fig. 4), *T-2H* model fits the first intermolecular peak of radial distribution curves significantly better than *T-1H* and MIX.

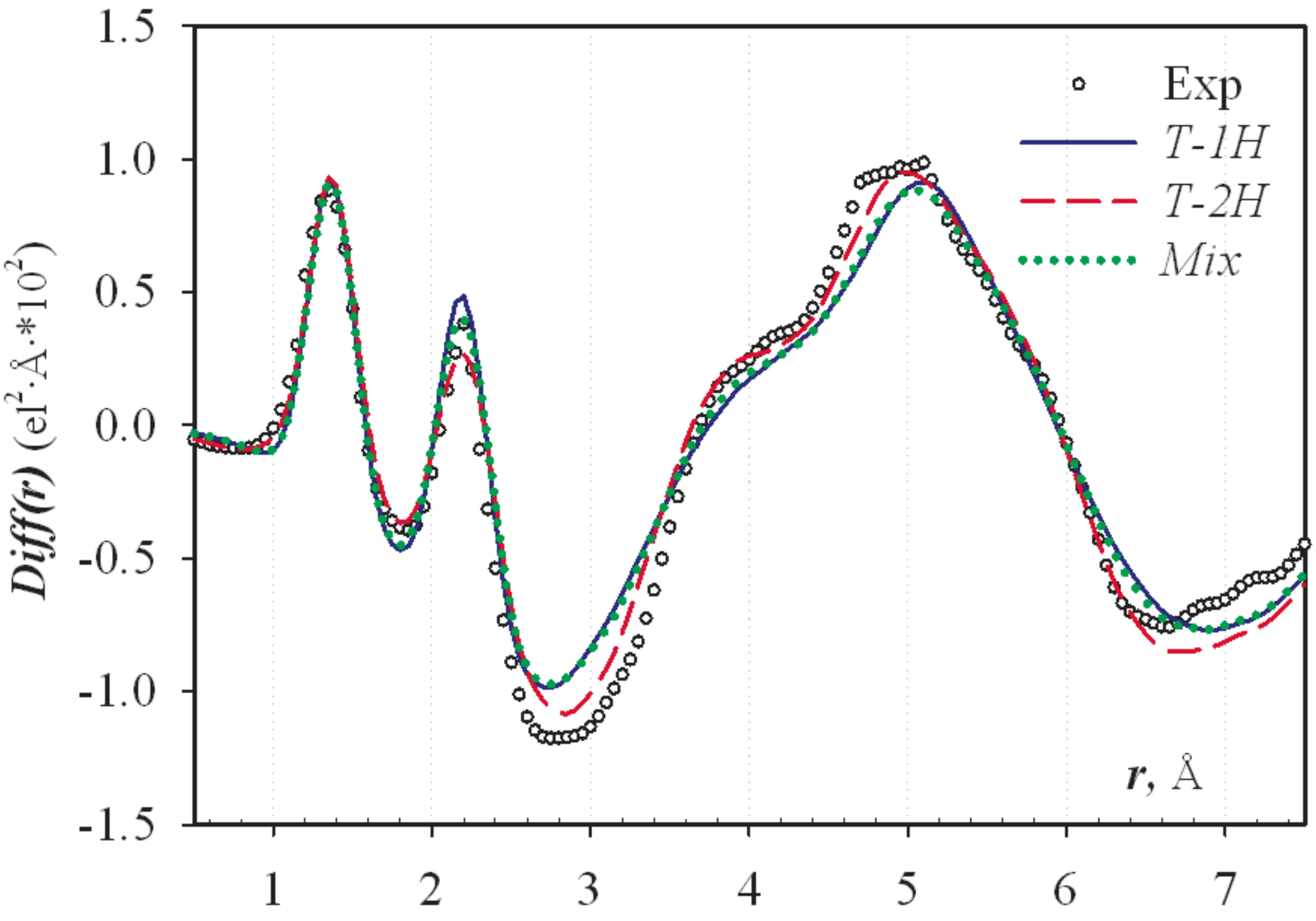

**Fig. 8 T1-H and T2-H MD models *vs* experimental**

The latter observation suggests that the geometrical arrangement of intermolecular interactions (e. g. H-bonds), which are responsible for the first molecular shell observed, is better represented in *T-2H* model. In any case, from these observations we can state that the liquid is certainly not composed of *T-1H* tautomer only, as it is indicated in specification data sheets of commercial products. The *T-2H* form gives a better agreement with experimental curves in most of the range observed, although the coexistence of both forms cannot be completely ruled out. The simple 1:1 mixture does not give any improvement of the curves with respect to *T-1H* and a quantitative assessment of the relative proportions of the two forms is hardly feasible with our X-Ray patterns.

More information about the intermolecular structure can be gathered from the analysis of the pair correlation functions (g(r)), calculated from molecular dynamics trajectory, bearing in mind that a direct comparison with experimental radial distribution functions is not possible, since contributions due to different atoms are not separable in X-Ray scattering patterns. Yet, the agreement between experimental and calculated patterns confirms the validity of the model in describing the structure of the system. From the analysis, it appears that the bulk structure of the liquid is governed by hydrogen bonding, with only a little contribution of molecular stacking (π-π interactions between aromatic systems). As it is evident from the intermolecular radial distribution curves of hydrogen bond donors and acceptors, reported in Fig. 9, strong interactions occur between donor *NH* groups of one molecule (N3 of *T-1H* and N2 of *T-2H*) and acceptor N: groups of another. The curves have an evident peak around the 3 Å region (3.13 *T-1H*; at shorter distance for *T-2H*, 3.03). The latter observation complies nicely with the better agreement observed for *T-2H* in Fig. 8, where it is evident that all the Diff(r) pattern is shifted at lower distances.

Each *1,2,3*-triazole may form a variable number of hydrogen bonds, (the peak integral over H N g(r) is 0.98, 0.68 for *T-1H* and, *T-2H* respectively) and it is evident from these figures that the *T-1H* tautomer has the biggest tendency to form hydrogen bonds; this fact may be explained in terms of repulsion between neighboring nitrogen atoms which is in turn counterbalanced by a highly polarized hydrogen atom. The visual inspection of trajectory frames and the pattern of centers of mass radial distribution function (dotted curve) suggests that molecular stacking is not very common (or the molecular planes are shifted or not parallel), since in that case we would observe an intense peak at distances shorter than 4 Å, while only a faint shoulder is observed in the centers of mass peak. Yet, the lack of stacking interactions could also be boosted by the inaccuracy of the force field used. It is well recognized, in fact, that most classical force fields (including GAFF/AMBER) fail in correctly describing π-stacking [31]. The periodicity of the COM radial distribution function (4.9; 8.75; 12.7) complies nicely with the oscillations observed in the experimental Diff(r) curve (Fig. 2), and suggests that outer molecules aggregate around a basic hydrogen-bonded dimer.

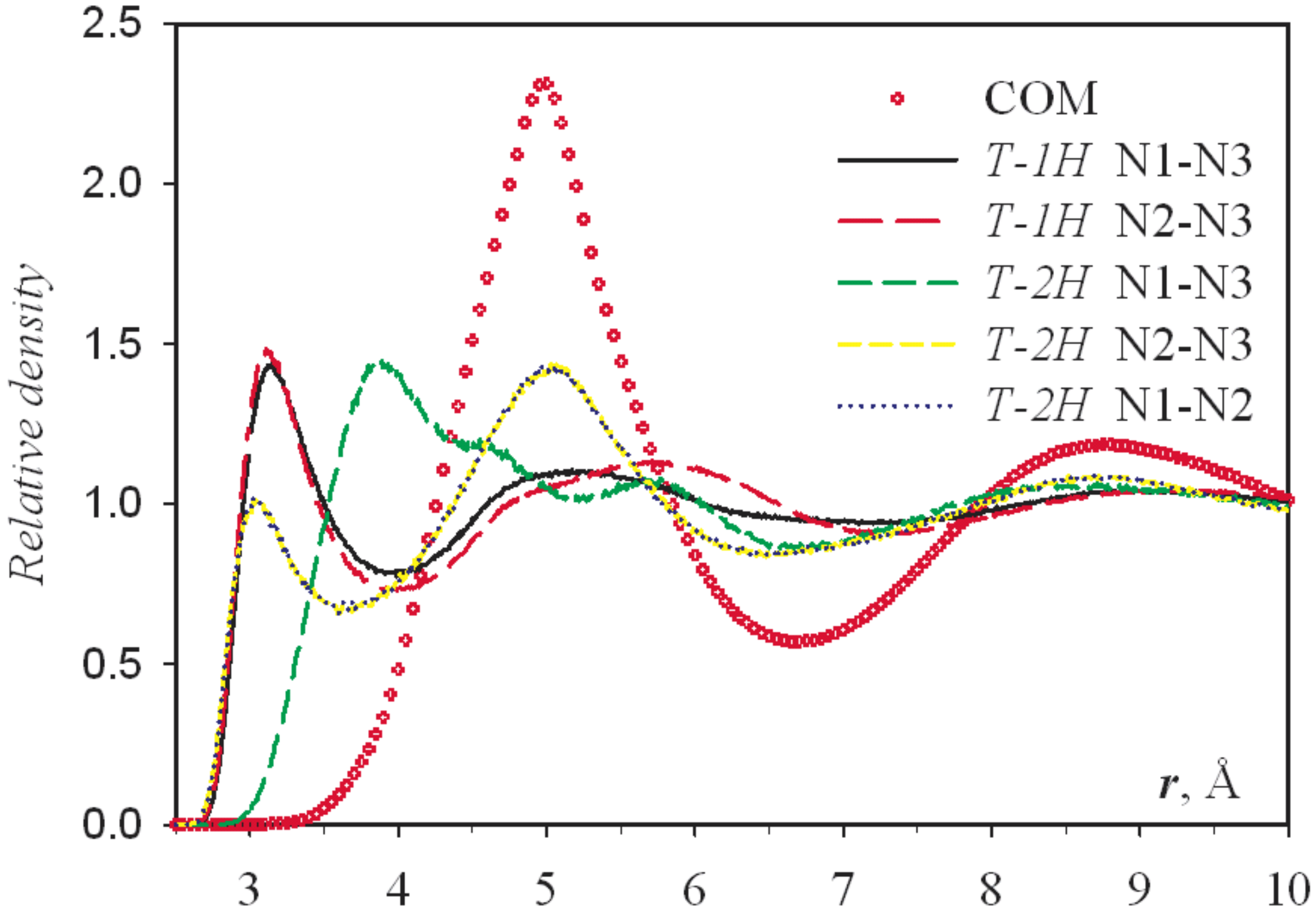

**Fig. 9 Pair distribution g(r): donor/acceptors atoms and center of mass**

## Computational and Infrared Spectroscopy Study

The computational study of these species was based upon the results of B3PW91 density functional (DFT) and Moeller-Plesset perturbation (MP) calculations. The equilibrium structures of the *T-1H* and *T-2H* tautomers, as well as of the transition state between them (the non-planar structure with the *NH* bond symmetrically placed between the *NN* bond) were optimized at DFT and MP2 levels employing several basis sets of type 6-311++G**, aug-cc-PVDZ, aug-cc-PVTZ and 6-311++G(3df,3pd). The three structures shortly described are shown in the following chart.

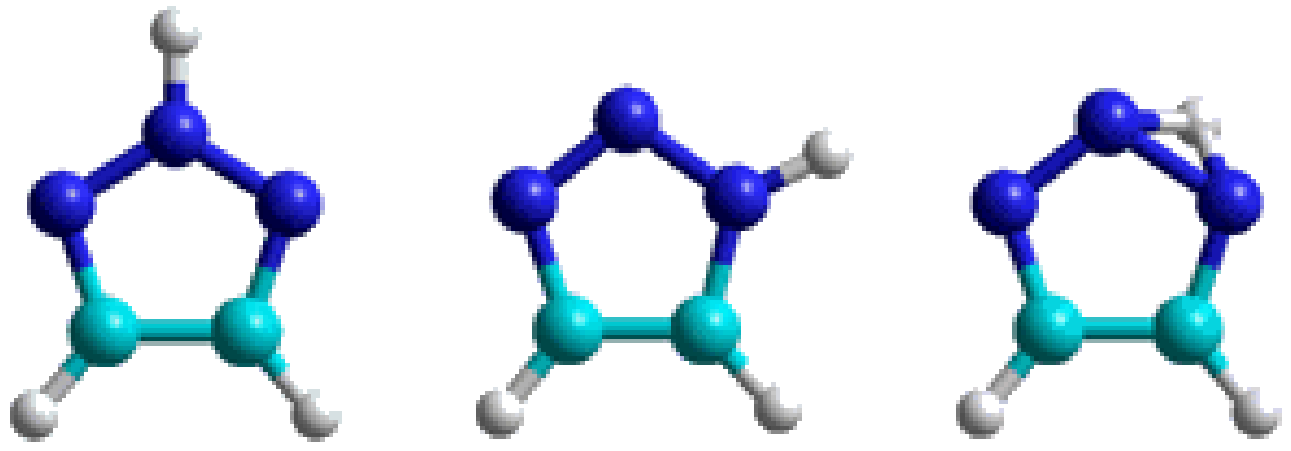

T2 T1 TS

T1

**Fig. 10 DFT *ab initio* geometries**

Each geometry optimization was followed by vibrational frequencies calculation with the multiple purpose of determining the nature of each stationary point (energy minimum for the stable structures and first-order saddle point for the transition state), zero-point and thermal corrections to the electronic energy values and, at last, the theoretical vibrational spectra of the two tautomers. The DFT calculations, as well as the single-point MP2 and MP4 energy ones, show that *T-2H* tautomer is always more stable than *T-1H*. A stability survey, accomplished through DFT, and particularly, from MP2 and MP4 calculations performed with the largest 6-311++G(3df,3pd) and aug-cc-PVTZ basis sets, suggests that the $C_{2v}$ tautomer is more stable than the less symmetric one by 18 ± 1 kJ mol$^{-1}$, in agreement with the conclusions reported by Toernkvist *et al*.[25] The transition state (TS) structure is lying 179 kJ mol$^{-1}$ higher in energy with regard to the less stable tautomer, similar to what was found by Rauhut [26]. Proton exchange from N1 to N2 through a planar path was found to produce a highly unstable model (the planar counterpart of TS) corresponding to a second order saddle point (above 200 kJ mol$^{-1}$ with respect to the lowest energy $C_{2v}$ structure). The very high tautomerization barrier is such to prevent gas-phase conversion of the $C_s$ symmetry tautomer into the most stable one. These considerations, strictly holding for gas-phase molecules, are strengthened by the FTIR matrix-isolation spectroscopy measurements carried out in argon matrix at 12 K, whose results are summarized in Table 2. The measured FTIR absorptions fit, at any computational level the DFT vibrational spectrum of the $C_{2v}$ symmetry tautomer, supporting the conclusions reported in the already cited microwave [4], as well as in a theoretical study [27] assessing that the largely predominant gas-phase tautomer is the $C_{2v}$ symmetry one. The assignment of the FTIR spectrum of the matrix-isolation investigation was based on several DFT and MP2 harmonic frequencies calculations performed with different basis sets, IR band intensities and potential energy distribution analysis. The theoretical and experimental results agree satisfactorily between them at any level of theory, irrespective of the choice of the B3LYP or B3PW91 functional and of the 6-311++G** or 6-311++G(3df, 3pd) basis set. This result is quite reassuring since, as described later in this work, the simulations of the infrared spectra of liquid *1,2,3*-triazole were performed with the smaller basis set (6-311++G**) because these calculations, owing to the size of the molecular aggregates considered for the purpose, could not be afforded with a larger basis.

Low-temperature aggregation of matrix-isolated *1,2,3*-triazole vapors was carried through warm-up cycles up to 35 K in order to produce *in situ* at least the smallest microcluster of the molecule, likely its dimer in argon matrix. Another aggregation study occurring at cryogenic temperature was accomplished by collecting the vapor with a large amount of helium gas onto the reflecting gold-coated copper target attached at the top of the cryogenic device. Helium gas acts as carrier gas to dilute and convey the vapor toward the reflecting support where extended molecular aggregation is induced by a series of thermal cycles (from 12 K to 35 K and back) yielding an extremely thin molecular layer. Both the FTIR spectra concerning with aggregation recorded at 12 K produced similar patterns revealing effects of bimolecular interaction of $C_{2v}$ type tautomers. Aggregation induces frequency shifts of some vibrations of the molecule (*NH* stretching and its *in-plane* and *out-of-plane* bending modes and the ring vibrations having high *NN* and *CN* stretching character) due to intermolecular hydrogen bonding. The simulation of the infrared spectra of the annealed argon-matrix spectrum, as well as of the molecular layer, was accomplished from DFT calculations of the spectra of a hydrogen-bonded cyclic-dimer of $C_{2h}$ symmetry and a cyclic-tetramer containing $C_{2v}$ type molecules (see Fig.11).

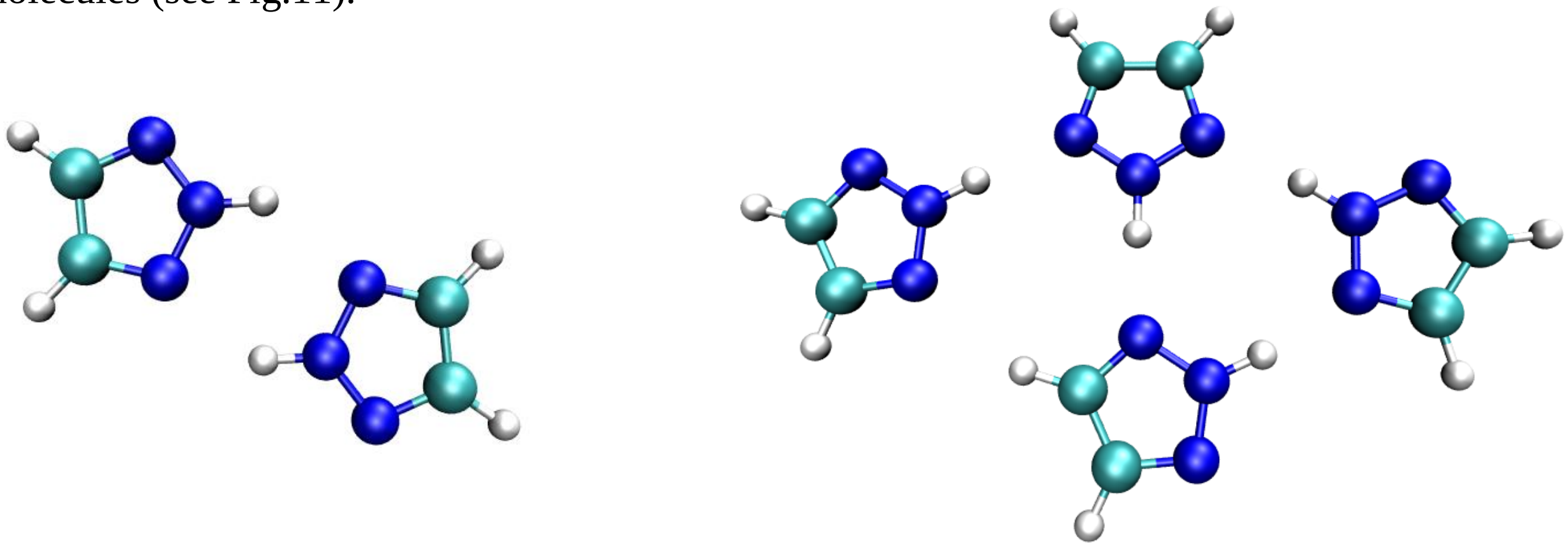

**Fig. 11 Dimer and Tetramer *ab initio* models (see text)**

The calculated frequencies of vibration of *1,2,3*-triazole molecule in the $C_{2h}$ symmetry dimer are reported in Table 3. Although the correspondence between calculated and measured vibrational wavenumbers is not always perfect, the simulation is actually satisfactory. In fact, one has to bear in mind that the simulation is performed within the harmonic approximation and calculated IR band intensity is an intriguing matter to be always handled with care. A quite different picture might occur in liquid phase where simultaneous presence of both the tautomers might be expected [28].

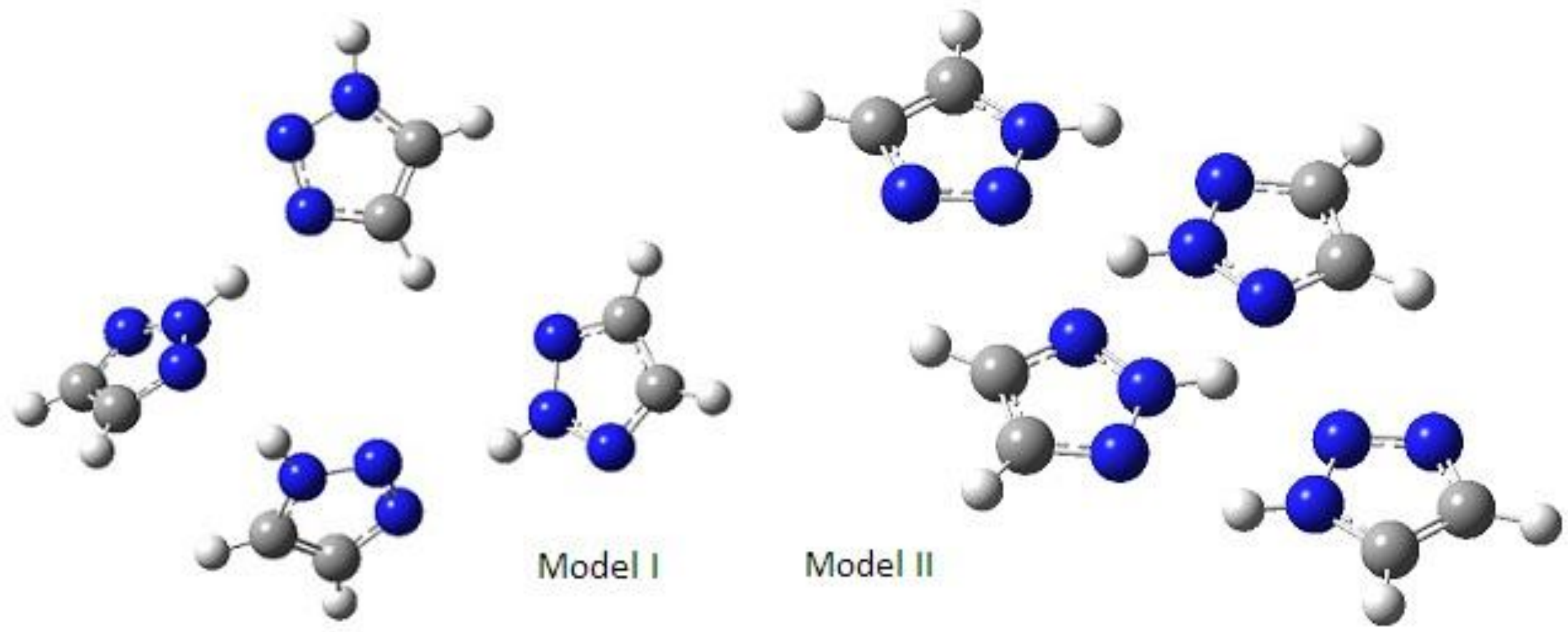

**Fig. 12 Supermolecular *ab initio* models used for IR spectrum calculation**

The FTIR spectrum of the liquid sample shows a number of bands consistent with a complex condensed phase. The observed bands do not appear particularly broad and the bands of the spectrum are well resolved allowing its interpretation reliably. This fact is quite important in the present study, as the infrared patterns of the two tautomers do not show too large differences. A series of theoretical simulations of the vibrational spectra were attempted at the DFT level to reach a reliable interpretative basis of the FTIR spectrum of the liquid sample. The simplest simulation was based upon the polarisable continuum model (PCM) approximation [29] considering each tautomer as a solute molecule placed in a continuum solvent having the dielectric constant of *1,2,3*-triazole ($\varepsilon$ =2.61) [30]. This simulation produces, with respect to the unperturbed vibrational force field, small frequency shifts and sometimes appreciable intensity changes for some bands of the two tautomers. Simulations of the spectrum of the liquid were accomplished taking into account more complex models consisting either of cyclic hydrogen-bonded dimers and tetramers. Further simulations considered "open" dimers and tetramers where interactions among molecules are weaker with respect to those of the cyclic aggregates. Within the limitations of the simplified structural units used to model the liquid phase, and bearing in mind that both the $C_{2v}$ and $C_s$ type tautomers are expected to be present in liquid phase, as assessed by EDXD and MD conclusions, the most satisfactory simulation of the FTIR spectrum, performed at the B3PW91/6-311++G** level, was provided by both the models (Model I and Model II) shown in Fig. 12. The summary of the measured and calculated FTIR spectra of the liquid compound is given in Table 4. Notwithstanding the FTIR spectra cannot assert whether Model I is preferred to Model II or *vice-versa*, they indicate that both the tautomers exist in appreciable amount in liquid phase, thus confirming the small energy difference and/or the lowering of the interconversion energy barrier in condensed phase. By the way, the computational study of the proton transfer reactions in *1,2,3*-triazole in gas phase and in aqueous solution do confirm a remarkable lowering of the energy barrier determined for some dimers of the molecule [26].

Table 2 - Summary of the FTIR study on *1,2,3*-triazole molecule in Ar matrix at 12K (vibrational frequencies are in $cm^{-1}$)

| **exp ($cm^{-1}$)** | **DFT Calculations** [(a)] | | | | | | **Assignment** [(b)] |
|---|---|---|---|---|---|---|---|
| | $C_{2v}$ | **I** | **II** | $C_s$ | **I** | **II** | |
| 3476 | $A_1$ | 3673 | 3674 | A’ | 3666 | 3669 | ν *NH* |
| | $A_1$ | 3282 | 3284 | A’ | 3295 | 3296 | ν *CH* |
| | $B_2$ | 3265 | 3267 | A’ | 3273 | 3274 | ν *CH* |
| | $B_2$ | 1553 | 1549 | A’ | 1550 | 1550 | $\delta_{ip}$ *NH* ν *ring* $\delta_{ip}$ *ring* |
| 1409 | $A_1$ | 1449 | 1450 | A’ | 1468 | 1468 | ν *ring* $\delta_{ip}$*CH* |
| | $B_2$ | 1424 | 1419 | A’ | 1384 | 1384 | ν *ring* $\delta_{ip}$ *NH* $\delta_{ip}$*CH* |
| | $A_1$ | 1308 | 1303 | A’ | 1313 | 1306 | ν *ring* $\delta_{ip}$ *CH* |
| 1235 | $B_2$ | 1280 | 1280 | A’ | 1193 | 1194 | ν *ring* |
| | $A_1$ | 1203 | 1203 | A’ | 1137 | 1139 | ν *ring* $\delta_{ip}$ *CH* |
| 1108 | $B_2$ | 1147 | 1141 | A’ | 1111 | 1110 | ν *ring* $\delta_{ip}$ *CH* |
| | $A_1$ | 1089 | 1090 | A’ | 1055 | 1059 | ν *ring* $\delta_{ip}$ ring |
| 961 | $A_1$ | 988 | 989 | A’ | 979 | 981 | $\delta_{ip}$ *ring* ν *ring* |
| 949 | $B_2$ | 966 | 965 | A’ | 955 | 955 | $\delta_{ip}$ *ring* |
| | $A_2$ | 891 | 898 | A’ | 854 | 879 | $\delta_{op}$*CH* |
| 826 | $B_1$ | 842 | 847 | A’’ | 776 | 783 | $\delta_{op}$ *CH* |
| 713 | $B_1$ | 729 | 731 | A’’ | 731 | 736 | $\delta_{op}$ *ring* |
| | $A_2$ | 679 | 682 | A’’ | 658 | 667 | $\delta_{op}$ *ring* |
| | $B_1$ | 582 | 580 | A’’ | 586 | 596 | $\delta_{op}$ *NH* |

(a) **I** stands for B3PW91/6-311++G** and **II** for B3PW91/6-311++G(3df,3pd) calculations

(b) ν (stretching), $\delta_{ip}$ (*in-plane* bending), $\delta_{op}$ (*out-of-plane* bending)

Table 3 - Summary of the FTIR study on the thin film from vapor condensation at 12K of gaseous *1,2,3*-triazole molecule (vibrational frequencies are in $cm^{-1}$)

| **exp.($cm^{-1}$)** | **cyclic dimer ($C_{2h}$) [b]** | **Assignment [a]** |
|---|---|---|
| 3370 | 3468 ($B_u$) | ν *NH* |
| 3144 | 3284 ($B_u$) | ν *CH* |
| 3025 | 3267 ($B_u$) | ν *CH* |
| | 1574 ($B_u$) | $\delta_{ip}$ *NH* ν *ring* $\delta_{ip}$ *ring* |
| | 1457 ($B_u$) | ν *ring* |
| 1413 | 1436 ($B_u$) | ν *ring* $\delta_{ip}$ *NH* $\delta_{ip}$*CH* |
| | 1307 ($B_u$) | ν *ring* $\delta_{ip}$ *CH* |
| 1254 | 1285 ($B_u$) | ν *ring* $\delta_{ip}$ *CH* |
| 1243 | 1210 ($B_u$) | ν *ring* $\delta_{ip}$ *CH* |
| 1124 | 1147 ($B_u$) | ν *ring* $\delta_{ip}$ *CH* |
| 1077 | 1095 ($B_u$) | ν *ring* $\delta_{ip}$ *ring* |
| 973 | 997 ($B_u$) | $\delta_{ip}$ *ring* ν *ring* |
| 959 | 973 ($B_u$) | $\delta_{ip}$ *ring* |
| | 884 ($A_u$) | $\delta_{op}$ *CH* |
| | 864 ($A_u$) | $\delta_{op}$ *CH* |
| | 826 ($A_u$) | $\delta_{op}$ *ring* |
| | 677 ($A_u$) | $\delta_{op}$ *ring* |
| | 668 ($A_u$) | $\delta_{op}$ *NH* |

(a) ν (stretching), $\delta_{ip}$ (*in-plane* bending), $\delta_{op}$ (*out-of-plane* bending)

(b) B3PW91/6-311++G** frequency calculations for the centro-symmetric cyclic dimer (two $C_{2v}$ *T-2H* type tautomers)

Table 4 - Summary of the FTIR study on liquid 1,2,3-triazole [(a)] (vibrational frequencies are in $cm^{-1}$)

| **exp ($cm^{-1}$)** | **Model I** | | **Model II** | |
|---|---|---|---|---|
| 1548 | 1572 | $\delta_{ip}$ *NH* $\nu$ *ring* $\delta_{ipr}$ r*ing* | 1605 | $\delta_{ip}$ *NH* |
| 1514 | 1554 | $\delta_{ip}$ *NH* $\nu$ *ring* $\delta_{ip}$*ring* $\delta_{ip}$ *CH* | 1570 | $\delta_{op}$ *CH* $\delta_{op}$ *NH* $\delta_{op}$ *CH* $\delta_{op}$ *NH* |
| 1501 | | | | |
| 1440 | 1486 | $\nu$ *ring*/ $\delta_{ip}$ *NH* $\delta_{ip}$ *CH* | 1475 | $\nu$ *ring* $\delta_{ip}$ *NH* $\delta_{ip}$ *CH* |
| | 1458 | $\nu$ *ring*/ $\delta_{ip}$ *CH* $\delta_{ip}$ *NH* | 1456 | $\nu$ *ring* $\delta_{ip}$ *CH* $\delta_{ip}$ *NH* |
| 1426 | 1442 | $\nu$ *ring* $\delta_{ip}$ *NH* $\delta_{ip}$*CH* | 1430 | $\nu$ *ring* $\delta_{ip}$ *NH* $\delta_{ip}$CH |
| 1426 | 1401 | $\delta_{ip}$ *NH* $\delta_{ip}$*CH* $\nu$ *ring* | 1411 | $\delta_{ip}$ *NH* $\delta_{ip}$*CH* $\nu$ *ring* |
| 1367 | 1316 | $\nu$ *ring* / $\delta_{ip}$ *CH* | 1321 | $\nu$ *ring* $\delta_{ip}$ *CH* |
| 1338 | 1306 | $\nu$ *ring* | 1302 | $\nu$ *ring* $\delta_{ip}$ *CH* |
| 1225 | 1291 | $\nu$ *ring* / $\delta_{ip}$ *NH* / $\delta_{ip}$ *CH* | 1288 | $\nu$ *ring* $\delta_{ip}$ *NH* $\delta_{ip}$ *CH* |
| | 1210 | $\nu$ *ring* | 1217 | $\nu$ *ring* |
| 1172 | 1205 | $\delta_{ip}$ *ring* / $\nu$ *ring* | 1206 | $\nu$ *ring* |
| | 1165 | $\nu$ *ring* /$\delta_{ip}$ *CH* | 1193 | $\nu$ *ring* $\delta_{ip}$ *NH* $\delta_{ip}$ *CH* |
| 1125 | 1151 | $\nu$ *ring* / $\delta_{ip}$ *CH* | 1160 | $\nu$ *ring* $\delta_{ip}$ *CH* |
| 1113 | 1121 | $\delta_{ip}$ *CH* $\nu$ *ring* | 1147 | $\delta_{ip}$ *CH* $\nu$ *ring* |
| 1093 | 1095 | $\delta_{ip}$ *CH* $\nu$ *ring* $\delta_{ip}$ *ring* | 1113 | $\delta_{ip}$ *CH* $\nu$ ring $\delta_{ip}$ *ring* |
| 1071 | 1081 | $\nu$ *ring* $\delta_{ip}$ *ring* $\delta_{ip}$ *CH* | 1068 | $\nu$ *ring* $\delta_{ip}$ *ring* $\delta_{ip}$ *CH* |
| 975 | 997 | $\delta_{ip}$ *ring* | 998 | $\delta_{ip}$ *ring* |
| | 977 | $\delta_{ip}$ *ring* $\delta_{ip}$ *CH* | 976 | $\delta_{ip}$ *ring* |
| 954 | 970 | $\delta_{ip}$ *ring* | 963 | $\delta_{ip}$ *ring* |
| | 965 | $\delta_{ip}$ *ring* | 927 | $\delta_{op}$ *NH* |
| 894 | 894 | $\delta_{op}$ *CH* | 885 | $\delta_{op}$ *CH* |
| 837 | 881 | $\delta_{op}$ *CH* $\delta_{op}$ *NH* | 885 | $\delta_{op}$ *CH* $\delta_{op}$ *NH* |
| 825 | 861 | $\delta_{op}$ *CH* $\delta_{op}$ *NH* | 857 | $\delta_{op}$ *CH* $\delta_{op}$ *NH* |
| | 842 | $\delta_{op}$ *NH* $\delta_{op}$ *CH* | 846 | $\delta_{op}$ *CH* $\delta_{op}$ *NH* |
| 790 | 819 | $\delta_{op}$ *NH* $\delta_{op}$ *CH* | 831 | $\delta_{op}$ *CH* |
| 758 | 793 | $\delta_{op}$ *CH* | 789 | $\delta_{op}$ *CH* $\delta_{op}$ *NH* |
| | 718 | $\delta_{op}$ *ring* | 728 | $\delta_{op}$ *ring* |
| 697 | 679 | $\delta_{op}$ *ring* | 694 | $\delta_{op}$ *ring* |
| 623 | 673 | $\delta_{op}$ *ring* $\delta_{op}$ *NH* | 677 | $\delta_{op}$ *ring* $\delta_{op}$ *NH* |
| 635 | 641 | $\delta_{op}$ *NH* | 663 | $\delta_{op}$ *NH* |

(a) B3PW91/6-311++G** frequency calculations for **Model I** and **Model II** (see Fig.11); $\nu$ (stretching), $\delta_{ip}$ (*in-plane* bending), $\delta_{op}$ (*out-of-plane* bending)

## Conclusion

In the present work, a combination of various experimental techniques and theoretical methods was employed to describe the hydrogen tautomerism occurring in *1,2,3*-triazole in different phases. FTIR measurements and *ab initio* calculations were used to gain a further insight into the presence of the *T-1H* and *T-2H* tautomers in gas phase, *i.e.* in argon matrix at 12 K, in a low-temperature film, as well as in liquid *1,2,3*-triazole, while Energy-Dispersive X-Ray Diffraction and molecular dynamics simulations were dedicated to study the phenomenon in liquid state. From our analysis it turns out that the *T-2H* form having the lowest *in-vacuo* energy, is the dominant form in gas phase while both the stable tautomers are the constituent species of the molecule in its liquid state. Diffraction measurements and MD simulations suggest that a mixture of *T-1H* and *T-2H* tautomers makes up the liquid. Though the theoretical X-Ray diffraction pattern calculated for a liquid phase constituted entirely by *T-2H* molecules is in better agreement with the experimental data, both form give a reasonable fit of the experimental structure function and radial distribution function.

## Acknowledgements

The authors thank the CASPUR consortium for the computing facilities used in this work (Standard 2012 grants std12-011 and std12-034). L. Gontrani acknowledges support from the FIRB "Futuro in Ricerca" research project (RBFR086BOQ_001, "Structure and dynamics of ionic liquids"